\documentclass{aa}  
\usepackage{booktabs}

\usepackage{graphicx}
            
\makeatletter
\newcommand*{\rom}[1]{\expandafter\@slowromancap\romannumeral #1@}
\makeatother
\usepackage{txfonts}
\usepackage{natbib}
\begin{document}

   \title{Properties and magnetic origins of solar S-bursts}

   \author{Brendan P. Clarke
          \inst{1, 4}
          \and
          Diana E. Morosan
          \inst{1, 2}
          \and
          Peter T. Gallagher
          \inst{1, 4}
          \and
          Vladimir V. Dorovskyy
          \inst{3}
          \and
          Alexander A. Konovalenko
          \inst{3}
          \and
          Eoin P. Carley
          \inst{1, 4}
          }

   \institute{School of Physics, Trinity College Dublin, Dublin 2, Ireland\\
              \email{clarkeb3@tcd.ie}
         \and
             Department of Physics, University of Helsinki, P.O. Box 64, Helsinki, Finland\\
             \email{diana.morosan@helsinki.fi}
         \and
             Institute of Radio Astronomy of NASU, Kharkov, Ukraine\\
             \email{dorovsky@ri.kharkov.ua}
         \and
             School of Cosmic Physics, Dublin Institute for Advanced Studies, Dublin, D02 XF85, Ireland\\
             }

   \date{Received July 24, 2018; accepted 6 January 2019}

  \abstract
   {Solar activity is often accompanied by solar radio emission, consisting of numerous types of solar radio bursts. At low frequencies ($<$100 MHz) radio bursts with short durations of milliseconds, such as solar S-bursts, have been identified. To date, their origin and many of their characteristics remain unclear.}
   {We report observations from the Ukrainian T-shaped Radio telescope, (UTR-2), and the LOw Frequency ARray (LOFAR) which give us new insight into their nature.}
   {Over 3000 S-bursts were observed on 9 July 2013 at frequencies of 17.4-83.1 MHz during a period of low solar activity. Leading models of S-burst generation were tested by analysing the spectral properties of S-bursts and estimating coronal magnetic field strengths.}
   {S-bursts were found to have short durations of 0.5-0.9 s. Multiple instruments were used to measure the dependence of drift rate on frequency which is represented by a power law with an index of $1.57$. For the first time, we show a linear relation between instantaneous bandwidth and frequency over a wide frequency band. The flux calibration and high sensitivity of UTR-2 enabled measurements of their fluxes, which yielded 11$\pm$3 solar flux units (1 SFU $\equiv10^4$ Jy). The source particle velocities of S-bursts were found to be $\sim$0.07 c. S-burst source heights were found to range from 1.3 $R_\odot$ to 2 $R_\odot$. Furthermore, a contemporary theoretical model of S-burst generation was used to conduct remote sensing of the coronal magnetic field at these heights which yielded values of 0.9-5.8 G.  Within error, these values are comparable to those predicted by various relations between magnetic field strength and height in the corona.}
   {}


   \maketitle
   
\setlength{\parskip}{0em}
\section{Introduction}
The Sun is an active star that regularly produces highly energetic phenomena such as flares and coronal mass ejections (CMEs). This activity is often accompanied by solar radio emission. In addition to the five standard classes of radio bursts (Types \rom{1}-\rom{5}), various types of radio bursts have been identified at metre and decametre wavelengths; these include metric spikes (\citealp{Benz1982a, Benz1996}), drifting spikes (\citealp{Elgaroy1979}), decametre spikes (\citealp{Melnik2014}), and supershort radio bursts (\citealp{Magdalenic2006}). These bursts usually have short durations of $\le$1 s and fine frequency structure, which may suggest that small-scale processes occur in the corona (\citealp{Morosan2015}). At longer wavelengths (>3 m), or frequencies <100 MHz, instrumental limitations have prevented detailed studies of fine structure radio bursts. However, using 32 broad-band dipoles, \cite{Ellis1969} was able to successfully identify a new type of solar radio burst at these frequencies that he called \textit{fast drift storm bursts}. These intriguing, low frequency bursts were investigated further by \cite{McConnell1982}, using the Llanherne Radio Telescope in Australia (4096 dipoles). McConnell renamed these \textit{solar S-bursts}, owing to their similarity to Jovian S-bursts: the S stands for short or storm. S-bursts appear as narrow tracks on a dynamic spectrum that usually drift from high to low frequencies, and in rarer cases, from low to high frequencies.

Figure 1 shows an example of a dynamic spectrum containing S-bursts, type \rom{3} bursts, and a type \rom{3b} burst. As our UTR-2 data\footnote{For access to this UTR-2 data, please contact Dr Vladimir Dorovskyy.} were flux calibrated, colour bars are included that indicate the flux values of the bursts.
\begin{figure*}[t!]
\centering
\includegraphics[width=15cm]{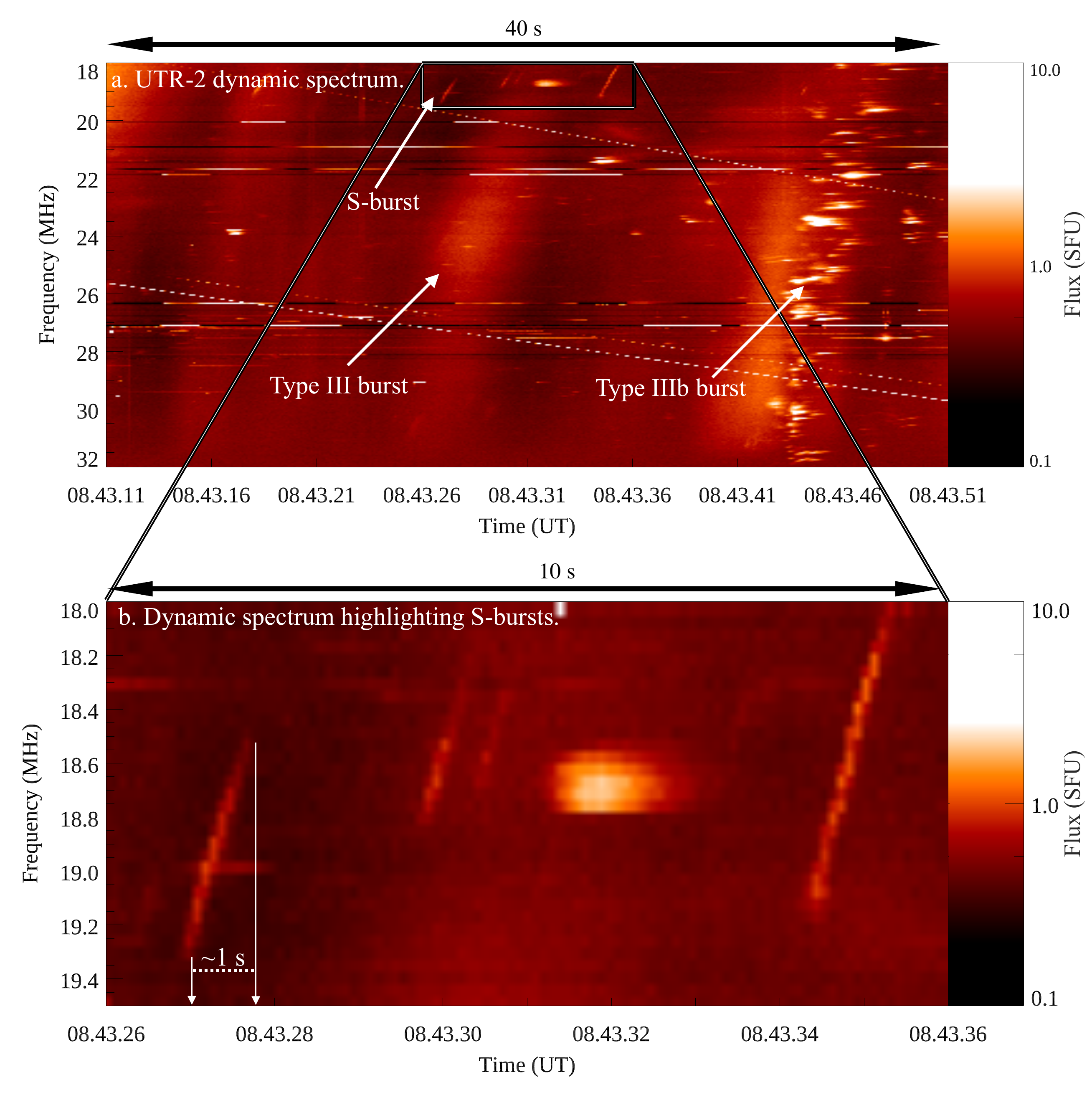}
\caption{Panel (a): Dynamic spectrum obtained from UTR-2 on 9 July 2013 containing examples of type \rom{3}, type \rom{3}b, and S-bursts. Several negatively drifting S-bursts are shown within the highlighted region between the times of 08:43:26 UT and 08:43:36 UT and the frequency range of 18-19.5 MHz. A type \rom{3} burst is also visible in this time frame. A type \rom{3}b burst is visible between the times of $08.43.41$ UT and $08.43.46$ UT. Panel (b): The highlighted region is shown in greater detail, indicating the short duration of the bursts ($<$1 s). Colour bars indicate the flux values of the bursts in solar flux units.}
\end{figure*}
S-bursts typically have a duration of $\le$1 s and have been observed to occur in a frequency range of 10-150 MHz, their number decreasing towards lower frequencies (\citealp{Ellis1969, McConnell1982, Melnik2010, Morosan2015, Dorovskyy2017}). S-bursts have short instantaneous bandwidths (frequency width at a fixed central time, $\Delta f$). \cite{Ellis1969} and \cite{McConnell1982} observed that the instantaneous bandwidths of S-bursts lie in a wide range. \cite{McConnell1982} observed values of 100 kHz at 30 MHz to 500 kHz at 80 MHz. Over a frequency band of 10-30 MHz, \cite{Melnik2010} found that the instantaneous  bandwidth of S-bursts increased linearly with frequency. S-bursts typically have a total bandwidth of approximately $2.5$ MHz (\citealp{McConnell1982, Morosan2015}) and in rare cases can reach highs of $10$-$20$ MHz, particularly in the decametre band (\citealp{Melnik2010, Morosan2015}). The full width half maximum (FWHM) duration (duration at fixed central frequency) was found to be <400 ms by \cite{Morosan2015}. S-bursts have low intensities, their fluxes rarely exceeding tens of SFU (\citealp{Melnik2010}).  They usually have negative drift rates that vary from $-7$ MHzs$^{-1}$ at $75$ MHz to $-0.6$ MHzs$^{-1}$ at $20$ MHz (\citealp{Morosan2015, McConnell1982}). S-bursts have been found to have a wide range of circular polarisation degrees from two to eight times that of type \rom{3} bursts (\citealp{Morosan2015}). S-bursts are associated with active regions and appear against the background of other types of solar radio activity such as type \rom{3} and \rom{3}b radio bursts (\citealp{Melnik2010}).

The exact emission mechanism behind the generation of S-bursts remains a topic of debate, however, they are believed to be emitted at the plasma frequency. In coherent emission mechanisms, electrons are accelerated in large numbers to produce electromagnetic radiation that is in phase. As large numbers of electrons are involved, the brightness temperature of the emission considerably exceeds the plasma temperature of the source region. Coherent emission mechanisms are identified in astrophysics in the form of solar radio bursts, stellar radio bursts, pulsar emissions, and planetary emissions. Examples of coherent emission mechanisms include plasma emission and electron cyclotron maser (ECM) emission (\citealp{Ginzburg1958, Wu1979}). Plasma emission requires a three wave interaction between a Langmuir wave, an ion acoustic wave, and an electromagnetic wave. Within the process, electron beams produced in active regions, as a result of events such as magnetic reconnection or shocks, propagate through the solar atmosphere. These electrons are accelerated differently depending upon their energy, causing the faster electrons to outpace the slower electrons. This results in an unstable velocity distribution known as a bump-on-tail distribution. This instability provides a source of free energy that gives rise to the growth of Langmuir waves, which may then scatter off ions to produce radio waves at the local plasma frequency. The second harmonic can also be produced in cases in which Langmuir waves propagate in opposite directions. For plasma emission in the solar corona, the frequency of the emitted plasma radiation is proportional to the local electron density, $N_{e}$. For fundamental emission, the plasma frequency, $f_{p}$ is \begin{equation} f_{p} = \frac{\omega_P}{2 \pi} = \frac{1}{2 \pi}\sqrt{\frac{N_e e^2}{m_e \epsilon_0}} \approx 9000 \sqrt{N_e} \end{equation} where $\omega_{P}$ is the angular frequency of the plasma radiation, $m_{e}$ is the electron mass, $e$ is the electron charge, and $\epsilon_{0}$ is the permittivity of free space. The electron density in the corona decreases as a function of height. Therefore, with the aid of an electron density model, the height of a radio source produced via plasma emission can be estimated. Electron density models can therefore provide approximate source heights at which radio sources occur. However, the electron density is lower over coronal holes (\citealp{Gallagher1999}) and higher above active regions (\citealp{Fludra1999a}). S-bursts are associated with active regions and so active region electron density and magnetic field models were used to estimate source heights and the magnetic fields at those heights for our analysis (\citealp{Morosan2015, McConnell1982, Melnik2010}). As solar S-bursts occur at low frequencies, their sources are expected to occur at high altitudes in the corona. Additionally, by assuming that S-bursts are produced as a result of the plasma emission mechanism, it is possible to estimate the velocity of the source electrons involved in the process via the frequency drift rate of the bursts. This can be achieved by using Equation 1 to find the electron density at the start time and end time of the burst. An electron density model can then be used to convert this information into a set of height versus time values which in turn can be used to infer the velocity of the exciter. \cite{McConnell1982} observed that S-bursts drift at a rate that is approximately three times slower than the drift rate of type \rom{3} radio bursts. This indicates that their source electron velocities are $\sim$0.1 c, which is approximately one-third of the velocity of typical type \rom{3} burst sources (\citealp{Morosan2015}). Several theories have been proposed to explain how S-bursts are generated. These are briefly introduced below.
\begin{figure*}[t!]
\centering
\includegraphics[width=17cm]{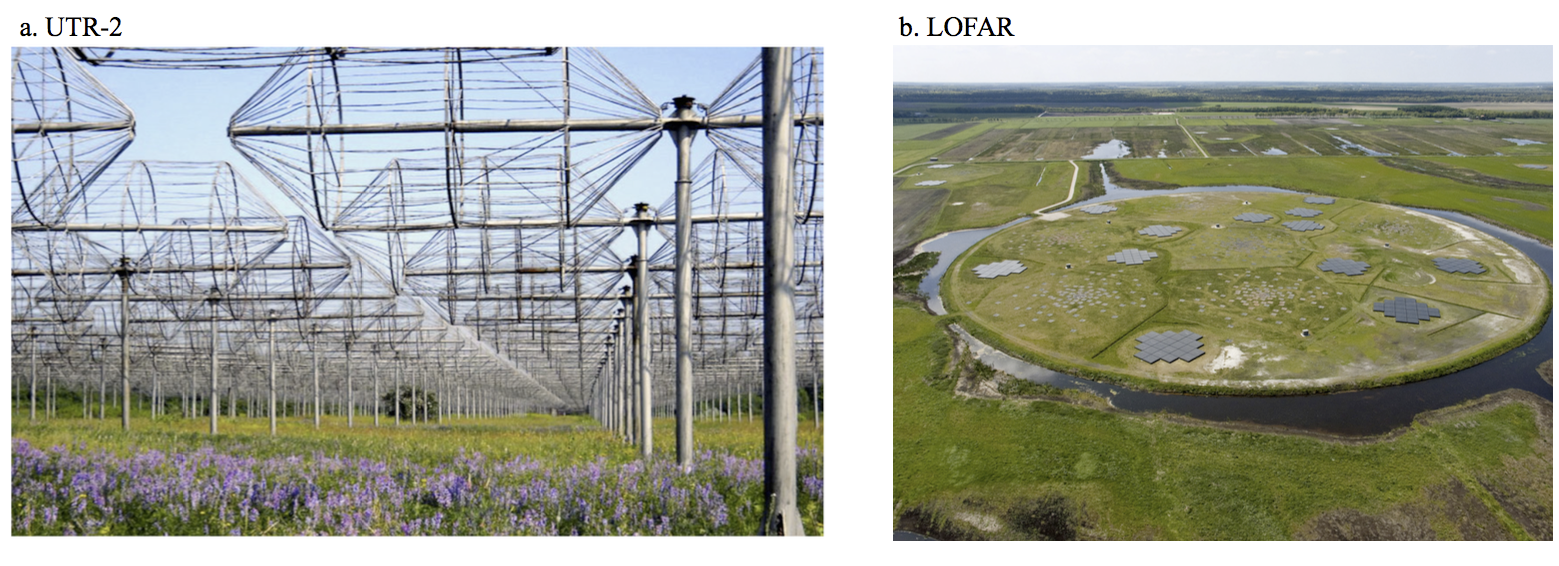}
\caption{Panel (a): Dipole elements that make up UTR-2. UTR-2 consists of 2040 of these dipoles. In the N-S direction, 1440 elements are spread over 240 rows, while 600 elements are spread over the E-W direction giving the telescope a total area of $\sim$1.4 $\times 10^{5}$ m$^{2}$ (\citealp{Konovalenko2016}). Panel (b): The heart of the LOFAR core, known as the Superterp (\citealp{Haarlem2013}). The large central island in the photograph contains 6 of the 24 core stations, each of which includes a field of 96 LBAs and two sub-stations of 24 HBA tiles.}
\end{figure*}

\cite{Melrose1982} argued that S-bursts are a variation of what are known as drift pairs. Drift pairs appear on a dynamic spectrum as two narrow parallel traces that have approximately the same frequency, but are separated in time by 1-2 s. Drift pairs are observed in cases in which two rays are reflected from a duct wall. This theory suggests that under certain conditions, when only a single ray is produced, we observe an S-burst at Earth. This theory has largely been dismissed becuase of the work of \cite{Melnik2010} and others that pointed out that drift pairs have considerably longer durations than S-bursts and no functional dependence of frequency on drift rate. Another model, put forward by \cite{Zaitsev1986}, suggests that plasma waves near the upper hybrid resonance frequency are excited owing to electrons moving in a slightly anisotropic plasma within a quasihomogeneous magnetic field that have velocities of 10 to 20 times the thermal velocity of the local electrons. These plasma waves are then scattered by thermal ions to produce electromagnetic radiation at the local plasma frequency, $f_{p}$. The negative drift rate in this model is attributed to group delay as the waves propagate outward from their point of origin. \cite{Melnik2010} argued that this model produces coronal inhomogeneity sizes and magnetic fields that are too high at heights corresponding to the local plasma frequency. Additionally, it implies a flatter drift rate dependence on frequency than follows from observations. 

More recently, \cite{Melnik2010} proposed a new mechanism of S-burst generation involving beams of particles (electrons or protons) in resonance with right-hand (RH) circularly polarised waves\footnote{The low frequency branch of this wave mode may sometimes be referred to as a fast magnetosonic wave \citep{Melnik2010}.}. This takes place against a background of Langmuir turbulence. The RH waves interact with the Langmuir waves to produce radio emission close to the local plasma frequency. This model attempts to account for all of the properties of S-bursts that the group observed, especially the following:
\begin{itemize}
  \item S-bursts are always observed against the background of other radio activity such as type \rom{3} and \rom{3}b bursts.
  \item S-burst sources move at velocities that are 5-9 times the thermal velocity of the local electrons.
  \item The instantaneous bandwidth of the S-bursts increases linearly with frequency. 
\end{itemize}

Using LOFAR, \cite{Morosan2015} observed that S-bursts may be associated with trans-equatorial loops that stretch to heights of $\sim$1.8 $R_\odot$. At such heights, it was noted that plasma emission is the most likely emission mechanism. It was observed that some of the properties of S-bursts, such as their relatively slow drift rates compared to type \rom{3} bursts, narrow bandwidths, and short durations are indicative of ECM emission. 
The conditions for ECM emission to occur are (a) a population inversion in the local electron distribution and (b) the frequency of emission must be greater than the local plasma frequency. For (b) to occur the following must be true:
\begin{equation}
\Omega_{e} = \frac{eB}{m_e} > \omega_{p} = 2 \pi f_{p}
\end{equation}
where $\Omega_{e}$ is the gyrofrequency and $B$ is the magnetic field at the source height of the bursts. Therefore a relatively strong magnetic field or low electron density is required for ECM emission to occur. \cite{Morosan2016a} found that this condition is only satisfied at heights of <$1.07$ $R_\odot$ and frequencies >$500$ MHz, indicating that ECM is not responsible for the generation of S-bursts. In this paper, we investigate the spectral characteristics of solar S-bursts and use these to investigate the feasibility of the \cite{Zaitsev1986} and \cite{Melnik2010} models outlined above. In Section 2, the instruments used for this study, UTR-2 and LOFAR, are introduced. An explanation of the observations and data analysis is also included in this section. In Section 3, we present our observations and analysis of the characteristics of S-bursts. In Section 4, we present a discussion of the proposed models of S-burst generation before concluding our findings in Section 5. 
\section{Instrumentation, observations, and data analysis}
\subsection{Instrumentation}
The Ukrainian T-shaped Radio Telescope \citep[UTR-2]{Konovalenko2016} and the LOw Frequency ARray \citep[LOFAR]{Haarlem2013} were used to observe S-bursts in this study. The UTR-2 telescope was built in the early 1970s in Ukraine. It is part of the interferometer known as the Ukrainian Radio Interferometer of NASU (URAN), which was completed in the final days of the Soviet Union. The telescope consists of 2040 array elements in two arms, north-south and east-west, which make up the T-configuration of the instrument. The north-south arm contains 1440 elements spread out over 240 rows while the east-west arm contains 600 elements spread out over 6 rows. The two arms of the telescope combine to give a total area of  $\sim$1.4 $\times 10^{5}$ m$^{2}$. The UTR-2 instrument operates within a frequency range 8-32 MHz with a frequency resolution of 4 kHz. This interferometer has an potential time resolution of 0.5 ms, however, for the observations included in this paper, the integration time was set to 100 ms. It is sensitive enough to detect signals that are $\geq$10 Jy and it has a dynamic range of 90 dB. It has an angular resolution of 40 arcminutes. Each element of the telescope is a broad-band, shunt type dipole composed of steel and zinc coated wires that are 8 mm in diameter. The dipoles in the form of two cylinders, as shown in Figure 2, orientated along the west-east line that have a diameter of 1.8 m, a length of 8 m, and a height above the ground of 3.5 m. Along the parallel, the dipoles are separated by 9 m while along the meridian they are separated by 7.5 m (\citealp{Braude1978}).

 The LOw Frequency ARray (LOFAR) is a large radio telescope network that was constructed by the Netherlands Institute for Radio Astronomy (ASTRON) in 2012. At the time of our observations (July 2013), the network consisted of $\sim$7000 antennas. Today, the network has grown and contains $\sim$20,000 antennas. The low band antennas (LBAs) of LOFAR operate at frequencies of 10-90 MHz, while its high band antennas (HBAs) operate at 110-240 MHz (\citealp{Haarlem2013}). These antennas are distributed over 24 core stations and 14 remote stations in the Netherlands, and 9 international stations across Europe. The 24 LOFAR core stations were used for the observations presented in this paper. One of the beam formed modes of LOFAR in the LBA frequency range was used to produce high time ($\sim$10 ms) and frequency (12.5 kHz) resolution dynamic spectra in order to study the spectral characteristics of the S-bursts detected on 9 July 2013 (\citealp{Stappers2011, Haarlem2013}). These LOFAR observations were first presented by \cite{Morosan2015} and are used in this work to complement the UTR-2 data analysis and to extend the analysis of S-bursts over a broader frequency range. More information about the LOFAR data analysis is provided in \cite{Morosan2015}.
 \begin{figure*}[t!]
\centering
\includegraphics[width=18cm]{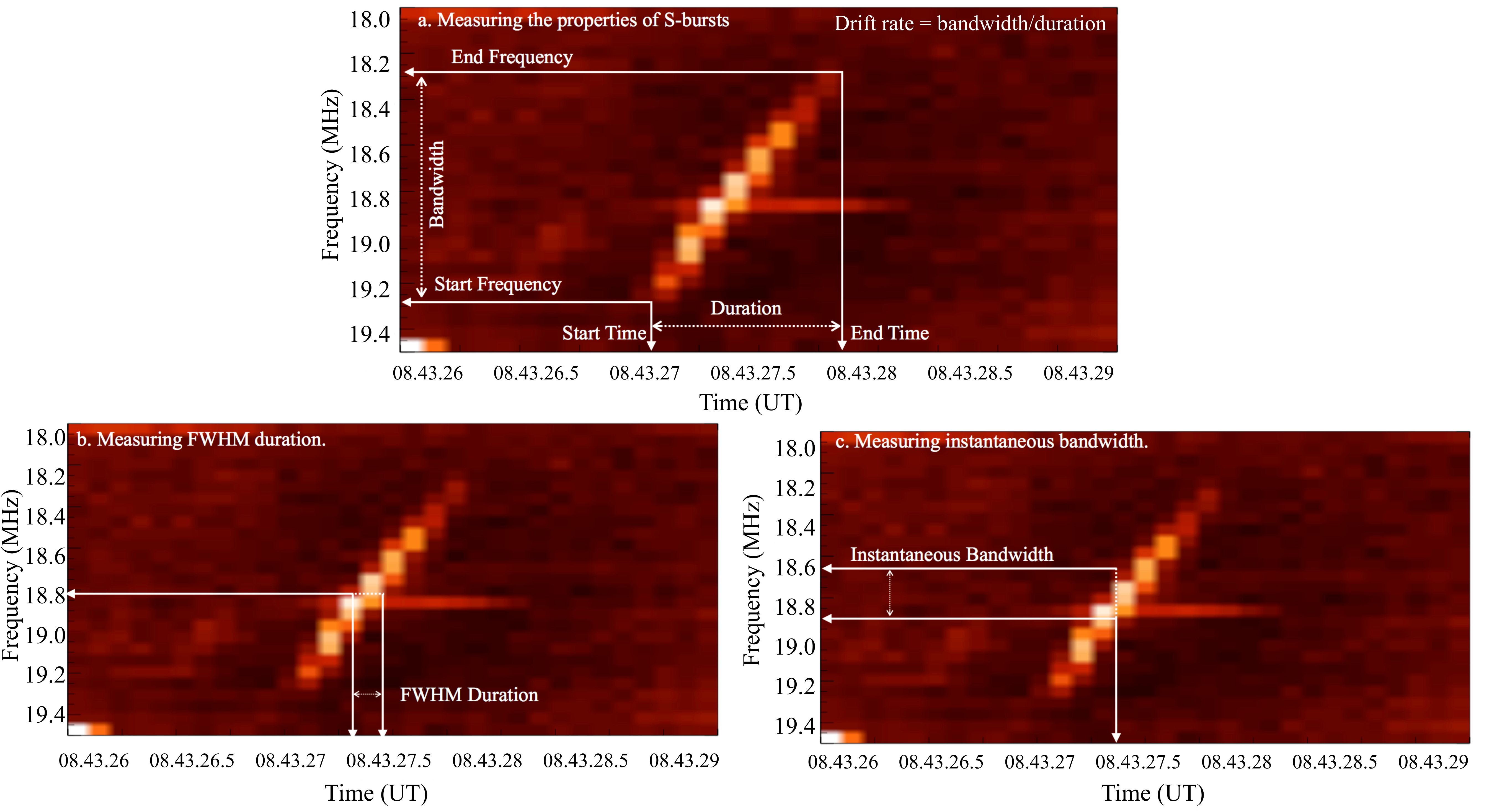}
\caption{Dynamic spectrum containing a typical, negatively drifting S-burst. Also shown is the method through which the various measured properties were recorded. Panel (a) shows the start/end time, start/end frequency, duration, bandwidth, and drift rate of the burst. Panel (b) shows the bursts FWHM duration. Panel (c) shows the bandwidth at a fixed central time (instantaneous bandwidth), sometimes referred to as the frequency width of the burst. The apparent stair-like appearance of the S-burst in this dynamic spectrum is an integration time effect and not an actual property of the burst.}
\end{figure*}
\begin{figure*}[t!]
\centering
\includegraphics[width=15cm]{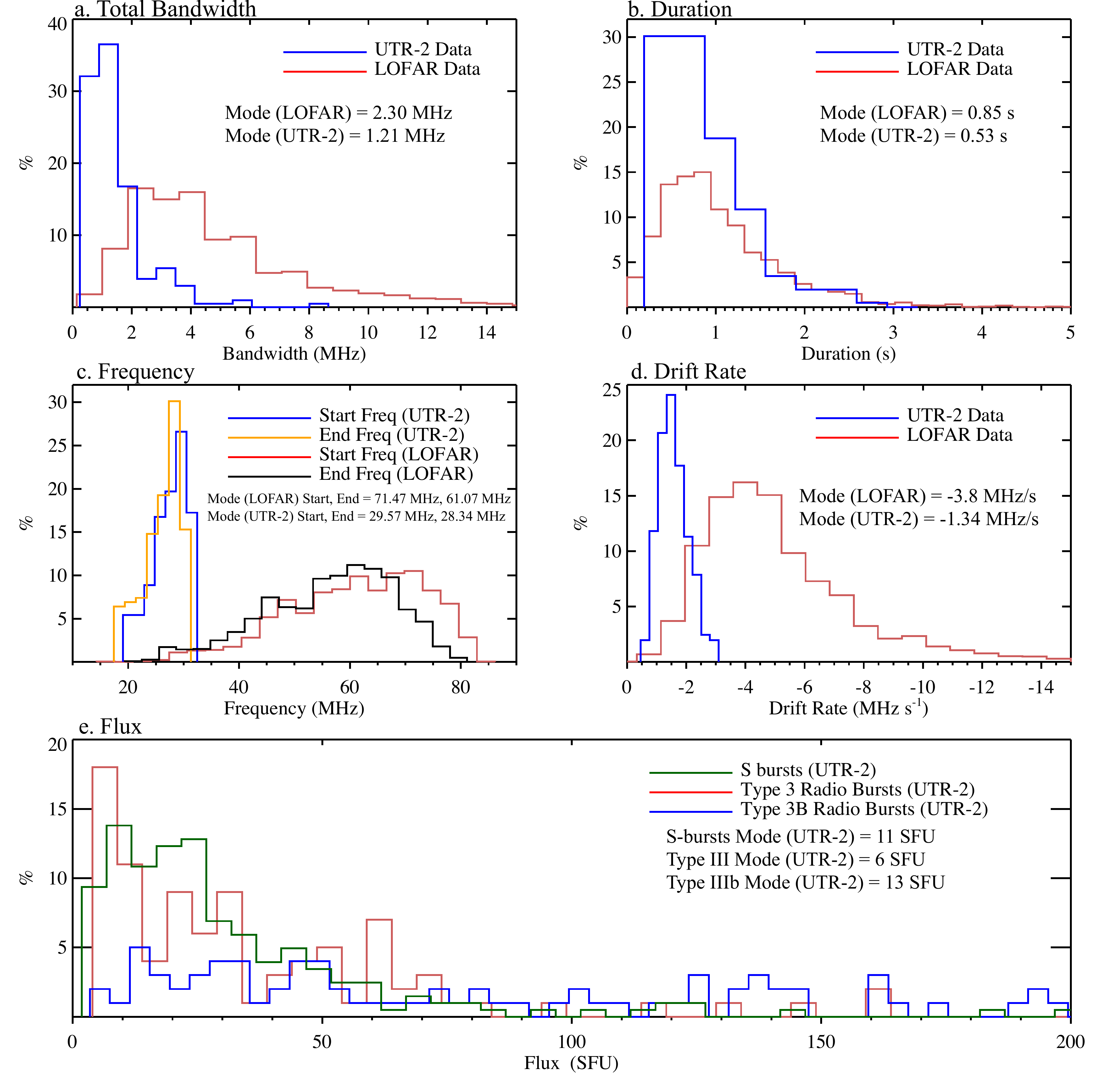}
\caption{S-burst property distributions from UTR-2 and LOFAR (\citealp{Morosan2015}). Panel (a) shows bandwidth; panel (b): duration; panel (c): start and end frequencies; panel (d): drift rate; panel (e): flux (only contains UTR-2 data; also includes data recorded for Type \rom{3} and Type \rom{3}b bursts). The mode values are given for each case.}
\end{figure*}
\subsection{Observations and data analysis}
\begin{figure*}[t!]
\centering
\includegraphics[width=12cm]{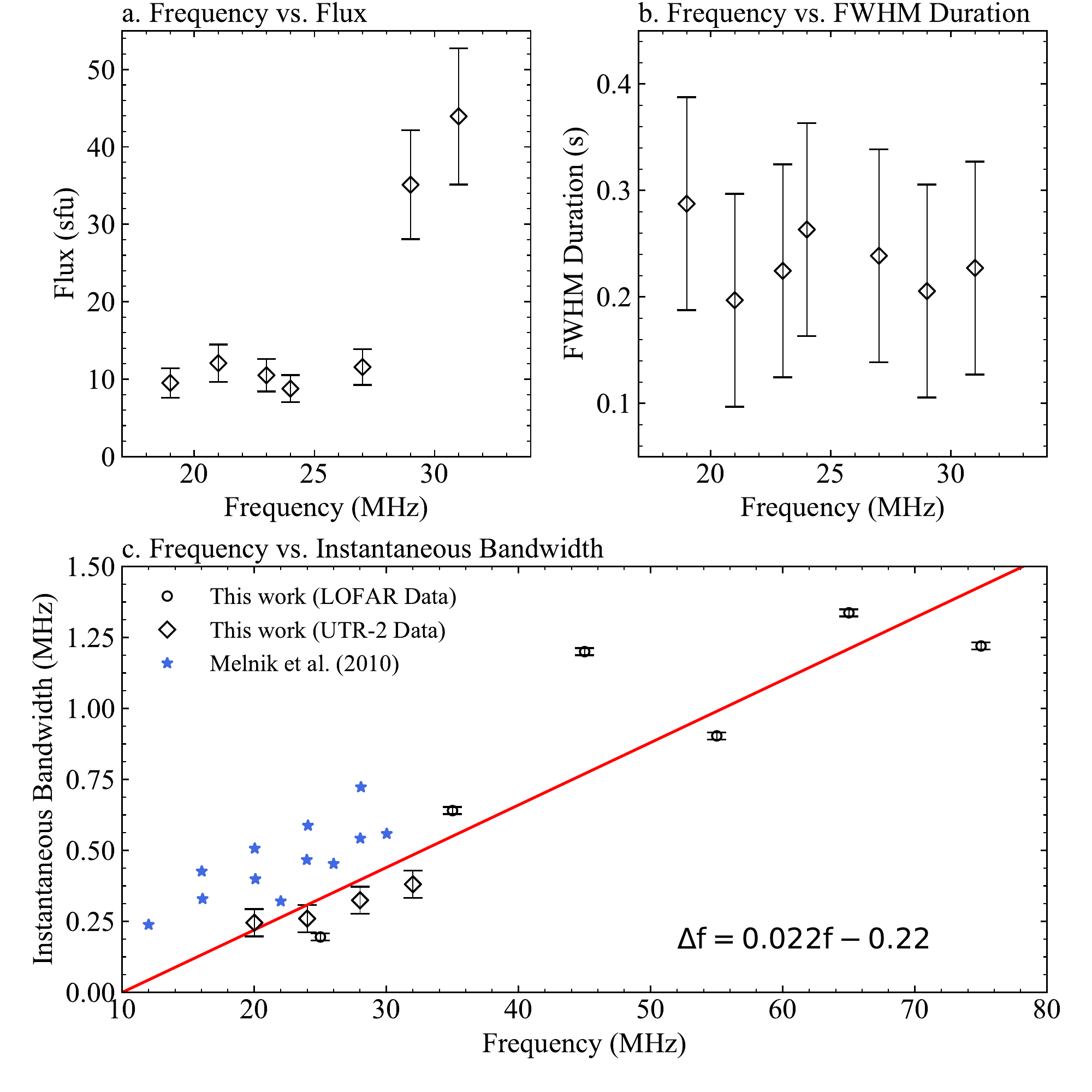}
\caption{Plots of frequency vs. flux and frequency vs. FWHM duration are shown in panels (a) and (b), respectively. No clear dependence was observed. In panel(c), a plot of frequency vs. instantaneous bandwidth is shown. Overplotted are the results of \cite{Melnik2010}. We observe a linear dependence between frequency and instantaneous bandwidth, confirming the same result observed by \cite{Melnik2010} over a broader frequency band.}
\end{figure*}

On 9 July 2013, 203 S-bursts were observed with UTR-2 and over 3000 S bursts were observed with LOFAR (\citealp{Morosan2015}). The observations were made between 05:34 and 14.30 UT. During this time, there was solar activity in the form of three C-class flares. Using UTR-2, 100 Type \rom{3} and \rom{3}b radio bursts were also observed during this period in order to compare their fluxes with that of the accompanying S-bursts. For each of the 203 events observed with UTR-2, the following properties of the bursts were recorded: start time, end time, duration, flux, start frequency, end frequency, bandwidth, drift rate, FWHM duration, and instantaneous bandwidth, as outlined in Figure 3. Figure 3a shows how the start frequency, end frequency, bandwidth, start time, end time, duration, and drift rate were recorded for a typical, negatively drifting burst.
Assuming the  bursts are produced via plasma emission, the total bandwidth provides information about how far through the solar atmosphere the source travelled while it was emitting. Figure 3b shows how the FWHM durations, which is duration at a fixed central frequency, of the bursts were recorded while Figure 3c shows how the instantaneous bandwidths, which is bandwidth at a fixed central time, $\Delta f$, were recorded. The instantaneous bandwidth provides information about the size of the source producing the emission, again assuming plasma emission (\citealp{Kontar2017}). The UTR-2 data were fully calibrated enabling flux measurements. The lighter colours indicate higher flux values as shown in the colour table in Figure 1. The low frequency band observations ($\sim$18-31 MHz) made with UTR-2 were compared against the higher and broader frequency band observations ($\sim$20-80 MHz) made with LOFAR (\citealp{Morosan2015}). The same properties that were measured with UTR-2 were measured with LOFAR. However, no flux measurements were conducted using LOFAR as the data were not flux calibrated. These measured properties were then analysed and compared against the assumptions made within the \cite{Melnik2010} and \cite{Zaitsev1986} models.

Assuming S-bursts are emitted at frequencies close to the local plasma frequency, it is possible to estimate their source heights. Using the active region coronal electron density models of \cite{NewkirkJr.1961}, \cite{newkirk1967}, Baumbach-Allen \citep{Aschwanden}, and \cite{Zucca2014a}, the observed S-bursts (frequencies of $\sim$18-80 MHz) are expected to have source heights of $\sim$1.3-2 $R_\odot$. The active region magnetic field model of \cite{McLean1978a} and a potential field source surface (PFSS) model, which provides an approximation of the coronal magnetic field up to 2.5 $R_\odot$ based on the observed photospheric field (\citealp{Schrijver2003}), can then be used to estimate the strength of the coronal magnetic field at these heights. In our analysis, we calculate the strength of the coronal magnetic field that the model of \cite{Melnik2010} predicts for each burst. We then estimate the source height for each burst and compare the strength of the magnetic field at each height, according to the active region magnetic field model of \cite{McLean1978a} and the PFSS model, with what the \cite{Melnik2010} model implies according to our data. 
\section{Results}
\subsection{S-burst properties}
All of the radio bursts that were observed with UTR-2 were identified manually due to the wide variation of intensity profiles and shapes contained within the data. The LOFAR data points were also identified manually (\citealp{Morosan2015}). Various properties of each burst were recorded. In Figure 4, the distributions of bandwidths, durations, start frequencies, end frequencies, and drift rates from both UTR-2 and LOFAR are shown. The flux distributions of S-bursts, Type \rom{3} bursts, and Type \rom{3}b bursts obtained from UTR-2 are also shown.

The S-bursts had an average total frequency bandwidth of $1.21\pm0.32$ MHz and $2.30\pm0.43$ MHz using the UTR-2 and LOFAR data, respectively. These results indicate that higher frequency S-bursts generally give rise to higher total bandwidths. This tells us that S-bursts sources with higher energies emit and travel over longer distances. Few bursts were observed to have total bandwidths greater than 10 MHz. After discarding unreliable data as a consequence of time resolution limitations, the instantaneous bandwidths of the S-bursts were found to range from 0.14-2.04 MHz. These results generally agree with the previous studies of \cite{McConnell1982}, \cite{Melnik2010}, and \cite{Morosan2015}. The mode duration was found to be $0.53\pm0.34$ s and $0.85\pm0.09$ using the UTR-2 and LOFAR data, respectively. As shown in Figure 4b, these distributions agree well within error despite the different frequency bands that UTR-2 and LOFAR can observe. This indicates that the duration of S-bursts is independent of the frequency of S-bursts. These values agree well with previous reports from  \cite{McConnell1982}, \cite{Melnik2010}, and \cite{Morosan2015}, most of whom observed S-bursts to have short durations of <1 s. The FWHM durations of the S-bursts were found to range from 0.1-0.6 s using the UTR-2 data and 0.02-0.4 s using the LOFAR data. These values also agree well. The integration time of our UTR-2 data (100 ms) most likely prevented detection of the lower values that were detected by LOFAR. The observed S-bursts had frequencies that ranged from 18.7-31.4 MHz (UTR-2) and 20.61-80.94 MHz (LOFAR). Figure 4c shows the distributions of start and end frequencies for each burst. The majority of S-bursts from both data sets had greater start frequencies than end frequencies illustrating how S-bursts generally start at higher frequencies and drift towards lower frequencies as they evolve. This tells us that S-burst sources generally travel from higher density regions lower in the solar atmosphere towards regions of lower density higher up.

S-burst fluxes are shown in Figure 4e in which they are compared to Type \rom{3} and Type \rom{3b} fluxes. Based on Figure 4e, Type \rom{3} bursts and S-bursts have similar flux distributions with S-bursts generally having higher fluxes. The distribution of flux values for Type \rom{3}b radio bursts is far broader than the other two types and there are many more examples of bursts with flux values >100 SFU. For these particular observations, the majority of Type \rom{3}b radio bursts have greater flux values than S-bursts or Type \rom{3} bursts. More generally, Type \rom{3} bursts can reach highs of $10^{6}$ SFU, Type \rom{3}b bursts are limited to thousands of SFU, and S bursts are rarely greater than 100 SFU (\citealp{Melnik2010, Reid2014}). 

Scatter plots of frequency versus flux and frequency versus FWHM duration are shown in Figures 5a and 5b, respectively. Figure 5c shows a plot of frequency versus instantaneous bandwidth. The included parameters were measured separately in several sub-bands. The sub-bands used in panels a and b were 18-20 MHz, 20-22 MHz, 22-24 MHz, 24-26 MHz, 26-28 MHz, 28-30 MHz and 30-32 MHz. In Figure 5c, the sub-bands used for the UTR-2 data are 18-22 MHz, 22-26 MHz, 26-30 MHz, and 30-34 MHz. For the LOFAR data, wider bands of 10-20 MHz, 20-30 MHz, 30-40 MHz, 40-50 MHz, 50-60 MHz, 60-70 MHz, and 70-80 MHz are used. The obtained values were linked to the central frequency of the corresponding sub-band. Overplotted  in Figure 5c are the results of \cite{Melnik2010}. No clear dependence was observed between frequency and flux or frequency and FWHM duration. Using the UTR-2 and LOFAR data collected on 9 July 2013, we observe a linear dependence between frequency and instantaneous bandwidth. The line of best fit is shown in red, which was found to have a slope of 0.022 and an intercept of -0.22 indicating a relation of the form
\begin{equation}
\Delta f = 0.022 f - 0.22. 
\end{equation} 
For this analysis, it was again ensured that because of time resolution limitations unreliable data were not included. This analysis confirms the same result observed by \cite{Melnik2010} over a broader frequency band using multiple instruments. 
\subsection{Drift rate}
An important property of radio bursts is how their frequency changes with time, known as their drift rate, $df/dt$. The relation between the drift rate and frequency of a radio burst can reveal information about the emission mechanism responsible for that burst. For example, if the bursts are generated at the local plasma frequency, we can estimate the velocities of the associated electron beams via this relation. The majority of S-bursts in this study had negative drift rates. However, there were some rare cases of positively drifting S-bursts, which agrees with previous authors (\citealp{Melnik2010, Morosan2015}). A plot of average frequency versus drift rate is shown in Figure 6. The dependence between drift rate and frequency may be represented by the following power law:
\begin{equation}
\frac{df}{dt} =-a f^{b},
\end{equation} 
where $a$ and $b$ are fitting parameters. The values of $a$ and $b$ derived from collective data obtained from UTR-2, LOFAR, and the Llanherne Radio Telescope are given in Table 1 providing a new analysis of this dependence from a multi-instrumental perspective. The values found from our collected UTR-2 data and those found in the previous works of \cite{McConnell1982}, \cite{Dorovskyy2017}, and \cite{MorosanGallagher2018} are also given. The values of $a$ and $b$ in Table 1 all agree well, which is reflected by the similarity of the fits shown in Figure 6. The slight discrepancy between the fits could be accounted for because of the different frequency ranges in which each set of observations were taken and because the measurements were taken on different dates. The date of the observations can have influence as the drift rate depends on the plasma properties, which may naturally vary from day to day. The distribution of drift rates for each S-burst is given in Figure 4d, where it is shown that the majority of the bursts drift at a rate of $-1.34\pm0.15$ MHz s$^{-1}$ and $-3.80\pm0.41$ MHz s$^{-1}$ using the UTR-2 and LOFAR data, respectively. This difference is to be expected due to the different frequency ranges of LOFAR and UTR-2. 
\begin{table}[h!]
\caption{Comparison of power law fit parameters.}
\begin{tabular}{@{}llll@{}}
\toprule
 Author(s) & $a$ & $b$\\ \midrule
\cite{McConnell1982}  & $0.0065\pm0.0006$  & $1.60\pm0.06$\\
\cite{Dorovskyy2017} & 0.0074 & 1.65\\
\cite{MorosanGallagher2018} & 0.0049 & 1.7\\ 
This Work (collective data) & 0.0084 & 1.57\\ \bottomrule
\end{tabular}
\end{table}
\section{Discussion}
\begin{figure}[t!]
\centering
\includegraphics[width=8.5cm]{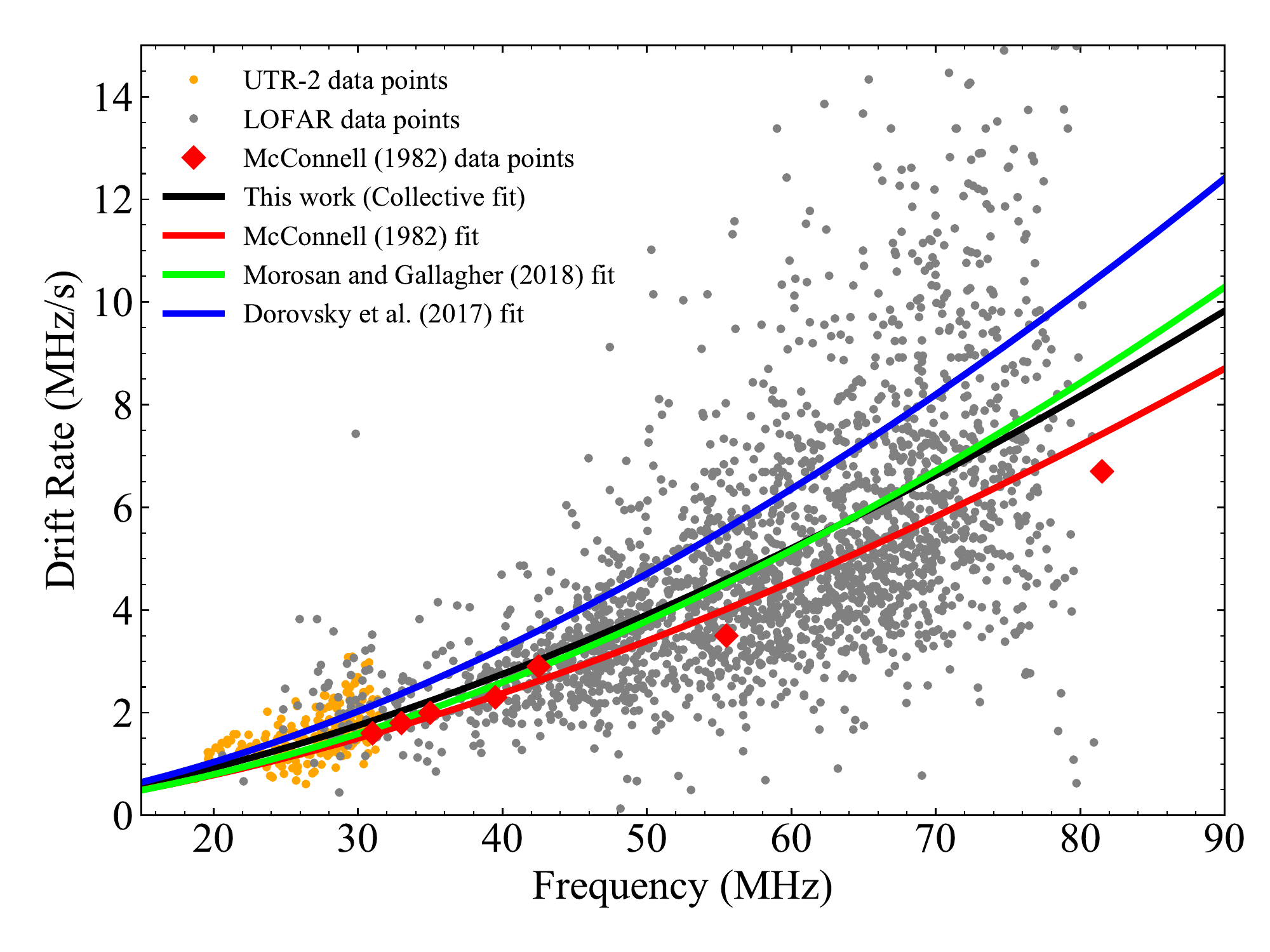}
\caption{Dependence of drift rate on frequency fitted according to power laws using coefficients $a$ and $b$. The fits derived from the observations of \cite{MorosanGallagher2018}, \cite{Dorovskyy2017}, and \cite{McConnell1982} are plotted in green, blue, and red, respectively. The black fit takes into account the orange (UTR-2 data), grey (LOFAR data from \cite{MorosanGallagher2018}), and red (data from \cite{McConnell1982}) points to provide, for the first time, a collective analysis of this dependence using multiple instruments.}
\end{figure}
The two models that we focus on in this discussion are the plasma emission models put forward by \cite{Melnik2010} and \cite{Zaitsev1986}. Each of these models assumes that the instantaneous bandwidth of S-bursts linearly increases with frequency. \cite{Melnik2010} showed this dependence over a narrow frequency band (10-30 MHz). Using UTR-2 and LOFAR data, our results can confirm this relation over a wide frequency band for the first time. 

The plasma emission model of \cite{Zaitsev1986} proposes that electron beams, with velocities that are ten to twenty times the thermal velocity of the local electrons, move through a slightly anisotropic plasma within a quasihomogeneous magnetic field. These beams excite plasma waves which are then scattered by thermal ions to produce electromagnetic radiation at the local plasma frequency, $f_{p}$. As outlined in the introduction, \cite{Melnik2010} pointed out that the model of \cite{Zaitsev1986} produces incorrect coronal inhomogeneity sizes and magnetic field strengths at heights corresponding to the local plasma frequency and a flatter drift rate dependence on frequency than follows from observations. Additionally, we note that the bursts modelled by \cite{Zaitsev1986} represent very narrow-band bursts (<1 MHz) that have central frequencies of $\sim$254 MHz and total bandwidths that are approximately equal to their instantaneous bandwidths. The model assumes a static source that attributes the frequency drift rate to the electromagnetic wave group delay. However, the S-bursts presented in this work display long lasting saber-shaped features which may extend in frequency by up to 12 MHz (\citealp{Melnik2010}). Additionally, their total bandwidths are much greater than their instantaneous bandwidths, indicating a dynamic source. Therefore, the model of \cite{Zaitsev1986} is unable to account for the characteristics of S-bursts that are commonly observed at decametre wavelengths. It is possible that the bursts modelled by \cite{Zaitsev1986} belong to a separate class of radio bursts that are produced by a different mechanism to the S-bursts we observe. Our observations of the spectral properties of S-bursts provide support for the core assumptions contained within the model of \cite{Melnik2010}. For example, the long-lasting sabre shaped structure of S-bursts and their appearance against the background of other types of radio activity were consistent throughout the data. A linear relation between frequency and instantaneous bandwidth was also found. Additionally, evidence was observed that S-bursts are produced by moving particles, as opposed to the stationary sources proposed by \cite{Zaitsev1986}.
 
The model of \cite{Melnik2010} suggests that S-burst sources are beams of electrons or protons which move at velocities that are five to nine times the thermal velocity of the local electrons. They propose that the merging of the RH waves with Langmuir waves gives rise to electromagnetic waves. Provided there is a sufficiently large angle between the k-vectors of these waves, the resulting emission has a frequency equal to the upper hybrid resonance frequency, $\omega_{UH}$, plus half the electron cyclotron frequency, $\Omega_{e} /2$. Given that the background Langmuir turbulence may have a wide angular spectrum, it is possible to find the range of angles between the Langmuir wave and RH wave k-vectors. Consequently, they show that
\begin{equation}
\frac{\Delta f}{f} = \frac{\Omega_{e}^{2}}{2\omega_{p}^{2}} \sin^{2}\theta,
\end{equation} 
where $f$ is the frequency and $\Delta f$ is the instantaneous bandwidth of the bursts. The value $\theta$ is the angle width of the Langmuir waves spectrum. This is a measure of the isotropy of the Langmuir waves involved in the interaction that produces the electromagnetic radiation. Rearranging equation 5 in terms of the magnetic field, $B$, at the source height of the bursts we get the following expression: 
\begin{equation}
B = \frac{\sqrt{8}\pi m_e}{e}f\sqrt{\frac{\Delta f}{f}}\frac{1}{sin \theta}.
\end{equation} 
\cite{Melnik2010} also derived a source velocity from the dispersion equation for RH waves propagating through the solar atmosphere, which is given below as
\begin{equation}
v_{s} = \frac{\Omega_{e}}{2 \omega_{p}}c.
\end{equation} 
This equation can be written as
\begin{equation}
v_{s} = \frac{eB}{m_{e}} \frac{1}{4 \pi f}c.
\end{equation} 

Given that the model assumes emission close to the plasma frequency, we can estimate the source heights for each burst. To do this, the active region coronal electron density models of \cite{Zucca2014a}, Baumbach-Allen \citep{Aschwanden}, \cite{NewkirkJr.1961}, and \cite{newkirk1967} were used. Figure 7a shows these electron density models, while Figure 7b shows how the plasma frequency varies with height for each model. Figure 7b also gives an example of how we estimated the source height of an S-burst with a frequency of $\sim$19 MHz. As shown, we calculated a minimum and maximum height and then used the average (dashed line) of these heights as our estimation. It is now possible to test this model further by inputting our data into equation 6 to calculate the source region magnetic field strengths that it predicts at specific heights above the solar surface. We then compared the strength of the magnetic field at each estimated height with the active region magnetic field model of \cite{McLean1978a} and a PFSS extrapolation. 
\begin{figure*}[t!]
\centering
\includegraphics[width=15cm]{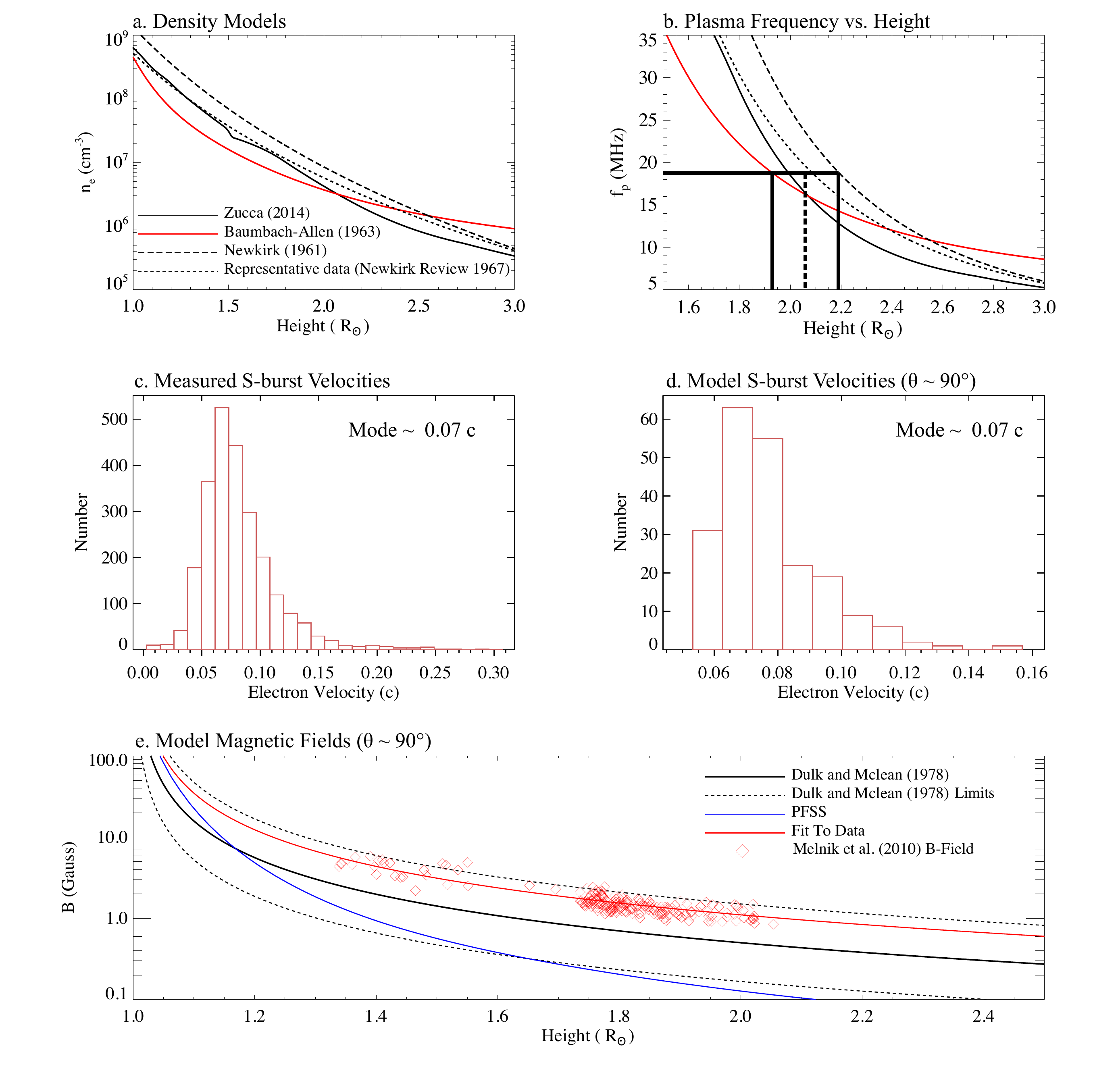}
\caption{Panel (a): Electron density vs. heliocentric height for the active region electron density models of \cite{Zucca2014a}, Baumbach-Allen \citep{Aschwanden}, \cite{NewkirkJr.1961}, and \cite{newkirk1967}. Panel (b): The plasma frequency vs. heliocentric height for each density model. An example of how we estimated the source height of an S-burst with a frequency of $\sim$19 MHz is shown. Panel (c): Distribution of measured source electron velocities. Panel (d): Source electron velocity distribution according to the model of \cite{Melnik2010} when using a value of $\theta=90^{\circ}$. Panel (e): The magnetic field strengths predicted by \cite{Melnik2010} at each of the source height estimates according to our data. In this case, a value of $\theta=90^{\circ}$ is used to allow for a source velocity of 0.07 c as implied by our observations.}
\end{figure*}

To calculate the magnetic field strength at the source height of the bursts according to the model of \cite{Melnik2010}, it is clear that a value of theta must be chosen. \cite{Melnik2010} found that this value must be between $50^{\circ}$-$90^{\circ}$ (partially or completely isotropic Langmuir turbulence) to satisfy the condition that S-burst sources move at velocities of $\sim$0.1 c. To constrain our choice of theta, we estimated the source electron velocities of all the S-bursts observed by UTR-2 and LOFAR. To do this, we calculated the source heights at the start and end of each burst via the density models, found the corresponding heights, and then calculated the velocities by dividing the distances travelled by the durations of the bursts. As shown in Figure 7c, the bursts were found to have a mode velocity of $\sim$0.07 c. A value of $\theta$ $\sim90^{\circ}$ produces a distribution of magnetic fields that results in a mode velocity of $\sim0.07$ c, according to equation 8, indicating isotropic Langmuir turbulence. Figure 7d shows the distribution of velocities calculated via equation 8. Figure 7e shows the magnetic field strengths predicted by the model of \cite{Melnik2010} at each of the source height estimates when we input our data. It was found that the magnetic field strengths ranged from 0.9-5.8 G between heights of $\sim$1.3-2 $R_\odot$. Overplotted is the active region magnetic field model of \cite{McLean1978a} and a PFSS extrapolation taken at a longitude corresponding to active region NOAA 11785, where the highest magnetic field strengths were observed at the time. The \cite{McLean1978a} model is accurate to within a factor of $\sim$3 between 1.02 $R_\odot$ and 10 $R_\odot$ and so the relevant error limits are included. The \cite{McLean1978a} model is given by the relation 
\begin{equation}
B = 0.5 \big[R/R_\odot - 1 \big]^{-1.5}.
\end{equation} 
As shown in Figure 7e, the data points lie between the standard and upper limit of the \cite{McLean1978a} model and appear to follow the same general form. A fit of the data was performed that allowed the coefficient (0.5) of Equation 9 to vary. A value of 1.1 was found and the corresponding fit is shown in red in Figure 7e. To verify the accuracy of the \cite{Melnik2010} model magnetic field strengths, we compared the magnetic field at 1.02 $R_\odot$ according to the PFSS extrapolation with the value found at the same height obtained via the extrapolation of the fitted data. The values were found to be $\sim$332 G and $\sim$389 G, respectively. These fields are in good agreement indicating that the fit to the data may provide us with the ability to conduct remote sensing of the coronal  magnetic field on the day of the observations. It is thought that PFSS extrapolations underestimate the coronal magnetic field at high altitudes of the solar atmosphere, however, they are thought to be more accurate closer to the surface. The slow decrease in the magnetic field according to the fitted data may be attributed to the higher than usual surface magnetic fields observed on 9 July 2013 of $\sim$1.7 kG in the active region NOAA 11785.

We note that S-bursts are very narrow-band phenomena that are characterised by localised pockets of emission that occur over a wide range of heights. To explain these pockets of emission, we propose that S-burst sources propagate along closed coronal loops. It is possible that the source particles are accelerated within the active region and then escape along a set of "stacked" coronal loops that stretch to higher and higher altitudes. The higher frequency emission would correspond to lower altitude loops, while the lower frequency emission would correspond to loops that stretch to higher altitudes. This would support our observations of reverse drift S-bursts and could explain their short durations and narrow frequency bands. High spatial resolution interferometric imaging is needed to confirm this idea.
\section{Conclusions}
Over 3000 S-bursts were observed in a frequency band of 18.7-83.1 MHz by UTR-2 and LOFAR on 9 July 2013. The S-bursts were found to have short durations of $\sim$0.5-0.9 s. The FWHM durations of the S-bursts ranged from 0.02-0.6 s. The S-bursts were observed to have a total frequency bandwidths of $1.21\pm0.32$ MHz and $2.30\pm0.43$ MHz using the UTR-2 and LOFAR data, respectively. The instantaneous bandwidths of the bursts ranged from 0.14-2.04 MHz. These results  agree well with those of previous authors (\citealp{McConnell1982, Melnik2010, Morosan2015, Dorovskyy2017}). A linear relation (slope of 0.022 and intercept of -0.22) between instantaneous bandwidth and frequency (central frequencies) was found, confirming the results of \cite{Melnik2010} over a wider frequency band. A functional dependence between the frequency and drift rate of S-bursts was observed using data from multiple instruments. This dependence was represented by a power law of the form $df/dt = -a f^{b}$, where $a$ was found to be 0.0084 and $b$ was found to be 1.57. These values are close to previously obtained results (\citealp{McConnell1982, Dorovskyy2017, MorosanGallagher2018}). The flux values of S-bursts, type \rom{3} radio bursts, and type \rom{3}b radio bursts were measured and compared. Type \rom{3}b radio bursts were found to have a broader flux distribution. No dependence of flux on frequency or flux on FWHM duration was observed. 

Leading theories of S-burst generation were investigated, particularly the models proposed by \cite{Melnik2010} and \cite{Zaitsev1986}. Several active region electron density models were used to estimate the heights at which the observed bursts would be generated ($\sim$1.3-2 $R_\odot$). It was noted that the model of \cite{Zaitsev1986} was unable to account for the properties of S-bursts that are commonly observed at decametre wavelengths. The source electron velocities of S-bursts were found to be $\sim$0.07 c. According to the model of \cite{Melnik2010}, it was found that the magnetic field strengths at the source heights of S-bursts ranged from 0.9-5.8 G. The model of \cite{Melnik2010} can enable us to conduct remote sensing of the coronal magnetic field. This model can account for the observed spectral properties of S-bursts and produced magnetic fields that are in good agreement with observations and coronal magnetic field models. However, a more detailed theoretical framework is needed to describe this mechanism in full. The study of S-bursts and their properties can help us to understand the physics of the coronal plasma. Higher time and frequency resolution imaging would enable us to study the spatial structure of S-bursts in more detail in order to further understand their nature.

\begin{acknowledgements}
      This work has been supported by the European Space Agency PRODEX Programme (BPC) and the Government of Ireland Studentship from the Irish Research Council (DEM). We acknowledge the Solar team from the Institute of Radio Astronomy of the National Academy of Sciences of Ukraine for conducting observations and providing the original data from the UTR-2 radio telescope. The team is composed of Dr. A. Stanislavsky, Dr. A. Koval, Dr. M. Shevchuk, and Dr. A. Boiko. This paper is based (in part) on data obtained from facilities of the International LOFAR Telescope (ILT) under project codes L158447, L158457 and L158521. LOFAR (\citealp{Haarlem2013}) is the Low Frequency Array designed and constructed by ASTRON. It has observing, data processing, and data storage facilities in several countries, which are owned by various parties (each with their own funding sources) and are collectively operated by the ILT foundation under a joint scientific policy. The ILT resources have benefitted from the following recent major funding sources: CNRS-INSU, Observatoire de Paris and Université d'Orléans, France; BMBF, MIWF-NRW, MPG, Germany; Science Foundation Ireland (SFI), Department of Business, Enterprise and Innovation (DBEI), Ireland; NWO, The Netherlands; The Science and Technology Facilities Council, UK7. E.P.C. is supported by the H2020 INFRADEV-1-2017 LOFAR4SW project no. 777442.
\end{acknowledgements}

\bibliographystyle{aa}
\bibliography{aanda.bib}

\end{document}